# Graphene-Quantum Dot Hybrid Photodetectors from 200 mm Wafer Scale Processing

*Sha Li,[1] Zhenxing Wang,[1,*] Bianca Robertz,[1] Daniel Neumaier,[1,2] Oihana Txoperena,[3] Arantxa Maestre,[3] Amaia Zurutuza,[3] Chris Bower,[4] Ashley Rushton,[4] Yinglin Liu,[4] Chris Harris,[4] Alexander Bessonov,[4] Surama Malik,[4] Mark Allen,[4] Ivonne Medina-Salazar,[4] Tapani Ryhänen,[5] Max C. Lemme[1,6,*]*

[1] AMO GmbH, Otto-Blumenthal-Str. 25, 52074 Aachen, Germany

[2] University of Wuppertal, Chair of Smart Sensor Systems, Lise-Meitner-Str. 13, 42119 Wuppertal, Germany

[3] Graphenea Semiconductor SLU. Donostia, Spain

[4] Emberion Limited, 150-151, Cambridge Science Park, Milton Road, Cambridge, CB4 0GN, UK

[5] Emberion Oy, Metsänneidonkuja 8, 02130 Espoo, Finland

[6] RWTH Aachen University, Chair of Electronic Devices, Otto-Blumenthal-Str. 2, 52074 Aachen, Germany

[*] Correspondence: wang@amo.de, lemme@amo.de

# Abstract

A 200 mm processing platform for the large-scale production of graphene field-effect transistor-quantum dot (GFET-QD) hybrid photodetectors is demonstrated. Comprehensive statistical analysis of electric data shows a high yield (96%) and low variation of the 200 mm scale fabrication. The GFET-QD devices deliver responsivities of $10^5$ - $10^6$ V/W in a wavelength range from 400 to 1800 nm, at up to 100 frames per second. Spectral sensitivity compares well to that obtained using similar GFET-QD photodetectors. The device concept enables gate-tunable suppression or enhancement of the photovoltage, which may be exploited for electric shutter operation by toggling between the signal capture and shutter states. The devices show good stability at a wide operation range and external quantum efficiency of 20% in the short-wavelength infrared range. Furthermore, an integration solution with complementary metal-oxide-semiconductor technology is presented to realize image-sensor-array chips and a proof-of-concept image system. This work demonstrates the potential for the volume manufacture of infrared photodetectors for a wide range of imaging applications.

# Introduction

Graphene possesses outstanding intrinsic properties such as monoatomic thickness, ultrahigh carrier mobility at room temperature, broadband optical absorption covering the far-infrared (IR) to ultraviolet range (UV), and an extremely high surface-to-volume ratio. As a result, graphene-based optoelectronic devices feature among the ideal candidates for photodetectors [1–3]. In fact, a variety of graphene-based photodetectors have been demonstrated in the lab that achieve outstanding performance on the device level, often outperforming their established semiconductor counterparts [4–10]. For example, graphene-based photodetectors have shown responsivities up to $10^8$ A/W and photoconductive gains up to $10^8$ electrons per photon in the IR regime [10]. Such detectors are potential candidates to extend the bandwidth of silicon (Si) complementary metal-oxide-semiconductor (CMOS) based photodetectors, which have limited sensitivity at wavelengths longer than 1.13 $\mu$m, i.e. with energy below the bandgap of Si. The high IR performance of those detectors was achieved through the hybrid integration of single-layer graphene (SLG) with colloidal quantum dots (QDs) as highly absorbing nanostructures [8,10]. Here, the wavelength of the light absorption can be tuned by selecting the size of the QDs. Furthermore, the monolithic integration of a CMOS integrated circuit with graphene field-effect transistor-QD (GFET-QD) photodetectors has also been demonstrated [8]. Although the signal output was several orders of magnitude lower than in the single-pixel reference devices, this early demonstration shows the potential for such devices to provide a low-cost route towards Si technology based visible to short-wavelength IR (VIS-SWIR) imagers.

A decisive remaining challenge towards industrial applications of such GFET-QD photodetectors is the availability of reproducible and wafer-scale "unit processes" [11–13], that will make graphene processing compatible with widely available conventional Si technology fabrication lines. This would be the preferred route toward the commercialization of two-dimensional (2D) materials in microelectronics, as it would require reasonable engineering efforts compared to the development of entirely new production lines from scratch. In addition, application-specific read-out integrated circuits (ROICs) are vital for optimizing the overall system performance, here in particular to reduce the noise level. The integration of graphene and 2D materials with CMOS circuits could thus take advantage of the well-established CMOS readout circuitry for signal processing purposes [8].

Great efforts have been directed at scaling up the synthesis, processing, device integration, and metrology of graphene and 2D materials up to the 300 mm wafer platform [12,11,14–18]. Despite the progress in processing technology, preserving the intrinsic performance of graphene at the device level remains challenging, in particular concerning wafer-scale uniformity and batch-to-batch reproducibility. We have identified the following unit processes that need to be established to realize scalable and high-quality production of high-performing GFET-QD photodetectors: high-quality growth and transfer, patterning with atomic precision, ohmic contacting, adhesion at the interface to graphene, design, and processing of the photosensitive stacks as well as hermetic semiconductor-packaging [17,19–23].

Here, we demonstrate a graphene fabrication process on a 200 mm wafer platform for large-scale production of GFET-QD photodetectors. This was realized by addressing several of the graphene wafer-scale integration challenges, including monolayer graphene chemical vapor deposition (CVD), transfer, and patterning that fulfill the requirements for imaging functionalization and large area deposition of the multilayer QD absorber material in an inert atmosphere, as well as methods for encapsulating the devices using thin film alumina ($Al_2O_3$) and hermetically sealed semiconductor packages.

## Results and Discussion

- **Device design and fabrication**

The tandem GFET-QD photodetector device architecture was designed based on graphene active regions covered with two different lead sulfide (PbS) QD layers as the light-sensitive absorber material. The devices include two contacts to the graphene and a buried gate electrode at the back that can be used to tune the device operating point. A schematic cross-section is shown in **Figure 1a**.

The working principle of the GFET-QDs tandem photodetector can be described based on the energy band diagram illustrated in **Figure 1b**. The semiconducting PbS QDs form a heterojunction with the graphene channel. The multi-layer absorber comprises a lightly doped or nearly intrinsic PbS QD layer with a narrow band-gap of $E_g = 0.64$ eV, which is capped by n-doped PbS QDs with a wide band-gap of $E_g = 1.25$ eV. If graphene is tuned to be p-doped, the heterojunction resembles a conventional *p-i-n* configuration that is commonly used in QD detectors and photovoltaics applications. The electronic band structure is thus designed to ensure a built-in electric field, which leads to hole transport towards the graphene layer upon optical excitation. The photo-excited charge carrier buildup in the active region upon illumination produces a photocurrent signal in the device, while the narrow bandgap of the PbS QDs determines the long-wavelength cutoff of the photo response. Such a GFET-QD device is shown in a scanning electron micrograph (SEM) in **Figure 1c**, which includes a local back gate, source and drain terminals, and the functionalization QD layer on top of the graphene channel.

The fabrication flow on 200 mm Si wafers is illustrated in **Figure 1d** and consists of the key processing modules of graphene transfer, patterning and contacting, and QD functionalization, as well as dicing and packaging. The as-fabricated 200 mm wafer can be seen in **Figure 1e** (left), and the best devices were packaged into hermetic semiconductor packaging as shown

**Figure 1e** (right) to allow further integration in a camera core. The process flow is based on photolithography and the details can be found in the Methods section.

A separate batch of quality control (QC)-wafers were fabricated with GFETs directly encapsulated by 60 nm $Al_2O_3$ without QD functionalization layers. This was done to monitor GFET processing and to evaluate the electrical parameters of graphene after device integration because QDs shift the Dirac Point by charge doping. In the following section, unless stated otherwise, the wafer scale statistics on graphene electric parameters were obtained from QC-wafers, whereas the photodetector parameters were obtained from the photodetector wafers where GFETs were fully functionalized with QDs.

- **Statistics on Wafer Scale**

We assessed the yield and reproducibility of the fabrication process as well as the suitability of graphene QD functionalization by a combination of non-invasive methods like optical microscopy and Raman-spectroscopy and electrical measurements of test structures. For each processing run, the quality of the monolayer graphene as well as the target parameters for yield, mobility, doping level, and contact resistance were quantified.

Defects and homogeneity across the area of transferred graphene were investigated by Raman spectroscopy and optical microscopy. **Figure S1** shows the inspection protocol, where characterization was carried out on dies covering the edge, middle edge, and center areas (**Figure S1a**). Each of the dies contains several devices, and two of the central devices (ID 5-3 and 5-4) were inspected at 20 and 100 magnifications. The image in **Figure S1d** shows a single graphene device, where the CVD graphene is visible in the central area. The brighter regions of the channel correspond to monolayer graphene, which covers most of the area, whereas the darker regions correspond to multilayer graphene and to wrinkles formed during the cooling of the growth process. This result shows most of the graphene to be in monolayer form, and of suitable quality for the targeted application.

**Figures S1e and f** show a Raman spectrum and Raman parameters extracted from a representative wafer, with the typical graphene fingerprints of the D band at approximately 1350 cm$^{-1}$, the G band at approximately 1582 cm$^{-1}$, and the 2D band at approximately 2700 cm$^{-1}$. The ratio between the intensity of the 2D and the G peak ($I_{2D}/I_G$) indicates the layer number of the graphene. The weak D peak in the Raman spectrum suggests a low number of defects, while the $I_{2D}/I_G$ ratio of approximately 3 confirms that the graphene is mostly a monolayer.

Automated current-voltage measurements were conducted to extract device characteristics and extract performance parameters. A photograph of the encapsulated GFETs on a 200 mm wafer is shown in the inset of **Figure 1f**). Out of the 648 devices randomly picked from the 200 mm wafer, a high yield of 96% has been achieved. Moreover, the fabrication flow was also found to be stable, with reasonable batch-to-batch reproducibility, confirmed by comparable fabrication yield (96% – 98%) from three fabrication runs as summarized in **Table S1**. The quantified analysis of the graphene electric parameters for 200 mm wafer is summarized in **Figures 1f-1i**, showing histograms of mobility, Dirac point ($V_{Dirac}$), hysteresis ($\Delta V_{Dirac}$), contact ($R_c$), and sheet resistance ($R_{sh}$) of the GFETs. Similarly, the analysis has been carried out on 150 mm wafers, with a high fabrication yield of 98.4%. A summary of graphene-on-wafer quality metrics on 150 mm can be found in **Table S2**, showing mobility, $V_{Dirac}$, $\Delta V_{Dirac}$, and $R_c$, $R_{sh}$ of the GFETs. Detailed analysis of statistics for both 200 mm as well as 150 mm wafers can be found in the supporting information.

- **Photodetection**

The detection performance of the photodetectors was measured and benchmarked with detectors that operate based on different detection mechanisms and are made from different materials and geometries. The key parameters of responsivity (R) and external quantum efficiency (EQE) are utilized in the following discussion. The responsivity of a photodetector is defined as the ratio of output photovoltage to incident light power on the photodetector. It is usually expressed as

$$R\ [A/W] = \frac{I_{ph}[A]}{P_{in}\ [W]} \qquad \text{Equation 1}$$

and

$$R\ [V/W] = \frac{V_{ph}[V]}{P_{in}\ [W]}, \qquad \text{Equation 2}$$

where $I_{ph}$ is the photocurrent, $V_{ph}$ is the photovoltage, and $P_{in}$ is the input optical power. This parameter is used to indicate the available output photocurrent or photovoltage of the photodetector for a given incident optical power at a certain wavelength.

EQE is the ratio between the number of electron-hole pairs with a contribution to the photocurrent and the number of incident photons. It can be expressed as

$$EQE = \frac{I_{ph}/e}{P_{in}/h\nu} = R \cdot \frac{hc}{e\lambda} \qquad \text{Equation 3}$$

where $e$ is the elementary charge, $h$ is Planck's constant, $\nu$ is the frequency of incident light, $c$ is the speed of incident light, and $\lambda$ is the wavelength of incident light.

Discrete GFET-QD devices were first measured using a probe station and parameter analyzer to obtain the transfer characteristics with a local back-gate voltage $V_{gs}$ = -5 to 5 V, as shown in the red curve in **Figure 2a**. Subsequently, the electrooptical response (*i.e.* photovoltage $V_{ph}$ as a function of $V_{gs}$) was measured within the same gate sweep range by pulsed illumination using a 520 nm laser diode at 0.5 Hz as shown in the black curve **Figure 2a.** Comparing the gate-dependent photovoltage at the graphene-QD junction with the transfer characteristics of the GFET-QD device shows a clear influence of the back gate bias, which tunes the depletion layer of the photosensitive QDs. This enables electrically controllable suppression or enhancement of the photovoltage, which can be exploited for realizing electric shutter operation by toggling the back gate between the signal capture and shutter states.

The performance of the GFET-QD photodetectors was further benchmarked by measuring the responsivity of a device as a function of chopping frequency and wavelength. In both cases, the gate bias was $V_{gs}$ = -1.0 V, with $V_{ds}$ = 0.5 V. **Figure 2b** plots the responsivity for chopping frequencies from 4 Hz to 100 Hz, at wavelengths of 520 nm and 1550 nm (in black and red, respectively). The responsivity of the detector decreases as the chopping frequency increases. In the measured frequency range, the responsivity was consistently slightly lower at the larger wavelength of 1550 nm compared to 520 nm, although the GFET-QD photodetectors exhibit high responsivities of $10^5$-$10^6$ V/W at 100 Hz for both wavelengths. The spectral response was further investigated by measuring the wavelength dependency of the photovoltage for $\lambda$ = 400 nm – 1800 nm, as shown in **Figure 2d**. The black and red curves represent responsivity spectra for 10 Hz and 100 Hz, respectively. Comparing the device response at the two frequencies, we note that, consistent with the results of **Figure 2b**, the higher frequency delivers lower responsivity. For $\lambda$ = 400 – 1800 nm, the responsivity remains rather high, which promises applications in a wider IR range.

The photoswitching behavior of the device was further evaluated by monitoring the time response under an on/off switched light source (**Figure 2c**). The device was measured upon pulsed light illumination at a wavelength of 1550 nm and a light intensity of P = 0.7 W/m². The corresponding photo response stabilizes approximately within 1s.

Lastly, EQE was measured as a figure of merit for the spectral range from $\lambda$ = 400 nm to 1800 nm to evaluate the conversion efficiency of the photodetectors (**Figure 2e**). The device maintains a high EQE of 100% for $\lambda$ = 400 to 500 nm. As $\lambda$ further increases up to 1000 nm, EQE decreases nearly linearly to 10% and stays below 20% up to 1800 nm, which can be explained by the position of the absorption peak for the PbS QDs, which in this case is around 1500-1520nm.

In summary, the GFET-QD photodetectors operate with a high responsivity of $10^5$ - $10^6$ V/W from 400-1800 nm, at up to 100 frames per second (corresponding to a response time of 10 ms) as well as wavelength sensitivity, comparable to that obtained using similar GFET-PbS QD photodetectors [24,25]. It also operates at a wide wavelength range (400 – 1800 nm), with particularly high energy efficiency of 100% in the 400 – 500 nm range due to the absorption peaks of the QDs.

Upon successfully developing the 200 mm wafer scale process for GFET-QD photodetectors, we realized image-sensor-array chips and proof-of-concept image systems. To that end, a hermetic semiconductor packaging solution was developed, which allows the integration with readout electronics and the creation of an imaging sensor and camera core. This involved designing and testing of a semiconductor package to prevent moisture and oxygen from degrading the sensor with the required number of input-output (IO) pins for connection to the application-specific integrated circuit (ASIC) on the parent wafer. Many different packaging options have been investigated with different suppliers and the first package closure trials have been performed to determine the effects of process parameters on the final device performance. The ROICs needed for data extraction from the sensor devices are not widely available as off-the-shelf items as the GFET-based sensors often require specific protocols and voltage levels to be used to obtain the best performance, so we developed a modular readout architecture for high-speed and low-noise applications being targeted. In first-generation single-pixel detectors, the ROIC was built on a separate printed circuit board (PCB) connected to the IR detector (**Figures 3a and 3b**). In later device generations, as in the case of the 512-pixel linear array detectors and video graphics array (VGA) imagers, the ROIC was integrated directly with the detector array sensor die. An example of a few-pixel GFET array in a quad flat no leads (QFN) package is shown in **Figure 3c**. In this case, the GFET-based sensors were fabricated directly on top of the planarized ASIC bare die. Once the fabrication steps were complete, a QFN package was used that had 64 pin-out connections. The detector die was bonded to the package, then the pads on the detector die were wire-bonded to the package pins before sealing in inert atmosphere. The photoresponse of a few-pixel array fabricated directly on the ASIC was measured through the ASIC at a back gate voltage of $V_{gs}$ = -1.4 V and -0.5 V, respectively, at a wavelength of 1550 nm and a light intensity of 0.7 W/m$^2$. The photocurrents in **Figures 3d and 3e** are plotted in arbitrary units that originate from digital values directly taken from the digital-to-analog converter (DAC) output of the ROIC.

## Conclusions

We have demonstrated the successful integration of QD-functionalized GFETs with CVD graphene on a 200 mm CMOS wafer platform, which showcases the route to volume manufacturing of IR photodetectors. Such sensors have a large number of imaging applications, from surveillance, search and rescue, and vehicle safety to improved sorting of food and food packaging to reduce environmental impact. The work illustrates the level of maturity of electro-optical devices based on monolayer CVD graphene as well as the current status regarding the main challenges associated with the growth, transfer, and patterning of graphene. The results further include the successful large-area deposition of multilayer QD absorber materials in inert atmosphere, including the methods to encapsulate the devices using thin film Al$_2$O$_3$ and hermetically sealed semiconductor packages. In summary, we have shown potential paths toward manufacturing graphene-based devices in the back end of the line and their packaging into a technology demonstrator for IR sensing.

# Methods

- **Fabrication of the Photodetector**

High-quality graphene was grown in a BM Pro 2×8" CVD furnace from Aixtron Ltd., using methane ($CH_4$) as precursor and copper (Cu, 200 mm × 200 mm foil) as the catalyst. Semi-dry transfer was carried out using sacrificial polymeric support placed on the graphene surface and the Cu foil was etched away using ammonium persulfate (APS) etchant. After rinsing the bottom surface of the graphene with DI water, it was dried and stamped on the target substrate using uniaxial pressure. Finally, the polymer was removed using solvents.

Local back-gated graphene GFETs were fabricated on 150 mm and 200 mm Si wafers with 90 nm silicon dioxide ($SiO_2$) using standard CMOS-compatible photolithography technology. Firstly, the titanium/palladium (Ti/Pd) local back gate with a thickness of 5/40 nm was evaporated onto the substrate, followed by a standard metal lift-off process. Then, a 75 nm thick $Al_2O_3$ layer was deposited hereafter using the atomic layer deposition (ALD) method to form the dielectric layer. Afterward, the bottom contact of 30 nm Pd was evaporated. After transfer, graphene was patterned using reactive ion etching (RIE) by oxygen plasma and then a final layer of 40 nm Pd was evaporated onto the substrate for top contacting.

The QD deposition was done by spin-coating a total of 12 layers of PbS QDs using layer-by-layer ligand replacement, in an inert atmosphere glovebox to minimize exposure to water and oxygen, followed by encapsulation using an aluminum oxide layer deposited using an ALD tool. There are 2 different sizes of QDs with 2 different ligands, making a 4-4-4 absorber type arrangement, with overall thickness of the order of 300 nm. **Figures S2a-d** show the 200 mm wafer before and after 4, 8, and 12-layer QD deposition, respectively. The absorber layer was then removed everywhere except on the graphene channels of the GFETs, using lithographic patterning and a solution-based etch process. The devices were finally encapsulated with $Al_2O_3$ using an ALD process. **Figure S2e** shows 2 quadrants of the 200 mm wafer after dicing, lithographic patterning, and ALD encapsulation. The absorber layers are well aligned on the graphene channels of the GFETs (close-up view in **Figure S2f**). After measurement using a probe station and laser diode to illuminate, the best devices were diced and wire bonded, and packaged into hermetic semiconductor packaging to allow further integration with readout electronics to create an imaging sensor and camera core.

- **Characterization**

Graphene on wafer samples were optically inspected under a Nikon LV100 microscope using a bright field, and Raman spectra were acquired using a Confocal Raman Microscope from WITec (Alpha 300) at a laser wavelength of 532 nm.

- **Electrical Measurements and Parameter Extraction**

All electrical measurements were performed at room temperature under ambient atmosphere using a two-point probe station and a semiconductor parameter analyzer (HP 4156B). Transfer characteristics $I_{ds}$-$V_{gs}$ of GFETs are measured with $V_{ds}$ of 100 mV and $V_{gs}$ in (-5, 5) V.

The TLM was used to extract the contact and sheet resistance. Local back-gated GFETs with channel length $L$ varying from from 9 to 89 $\mu$m in a step of 10 $\mu$m and a channel width $W$ of 19 $\mu$m are fabricated. The total resistance $R_t$ consists of channel resistance from graphene and contact resistance from graphene/metal junction can be expressed as $R_t(L) = R_{sh}L/W + 2R_c/W$. The total resistance ($R_t$) of each neighboring contact pair was measured as a function of back-gate voltage ($V_{gs}$), with a $V_{ds}$ of 100 mV. All devices show the typical behavior of GFETs, where the resistance maximum indicates the charge neutrality point ($V_{Dirac}$). By linear fitting $R_t$ of GFETs with different $L$ under the same net gate voltage ($V_{gs}$-$V_{Dirac}$), $R_{sh}$ and $R_c$ at this gate voltage can be extracted from the slope and intercept respectively. Gate voltage-dependent $R_{sh}$ and $R_c$ curves can also be obtained.

Mobility was extracted by the field-effect mobility model, called direct DTM [26–28]. Using the gate voltage-dependent transconductance of the GFETs, the field-effect mobility $\mu$ is calculated as following $\mu = g_m L/(W V_{ds} C_{gs})$, where transconductance $g_m = \partial I_{ds}/\partial V_{gs}$, graphene channel length $L$ and width $W$, drain voltage $V_{ds} = 100$ mV; gate capacitance $C_{gs} = \varepsilon\varepsilon_0/t_{ox}$, with $t_{ox} = 90$ nm, $\varepsilon = 3.9$ (SiO$_2$) and $\varepsilon_0 = 8.85 \times 10^{-12}$ F/m.

The electro-optical response of the GFETs was measured using an MPI probe station and Keithley BF1500 parameter analyzer to obtain the transfer curves of the GFET-QDs as a function of back-gate voltage, followed by pulsed illumination using 520 nm and 1550 nm laser diodes at 0.5 Hz. The photovoltage as well as the voltage response was calculated based on the resistance between source and drain of the GFET. More details can be found in supporting information.

## Acknowledgments

This work has received funding from the German Ministry of Education and Research (BMBF) under the project GIMMIK (03XP0210), and from the European Union's Horizon 2020 research and innovation program under the grant agreements 820591 (G-IMAGER), 881603 (Graphene Flagship Core3) and 952792 (2D Experimental Pilot Line).

## Author contributions

Z.W., D.N., A.Z., C.B., T.R., and M.L. conceptualized the idea. S.L. and B.R. performed the experiments of wafer scale device fabrication, as well as the electrical characterization. O.T. and A.M. performed wafer scale graphene growth and transfer. C.W., A.R., Y.L., C.H., A.B., S.M., M.A., and I.M.-S. performed the design of the device, the experimental work of quantum dots coating as well as the optical characterization. Z.W., A.Z., T.R., and M.L. supervised the project. All authors contributed to drafting and revising the manuscript.

## Competing interests

The authors have no competing interests.

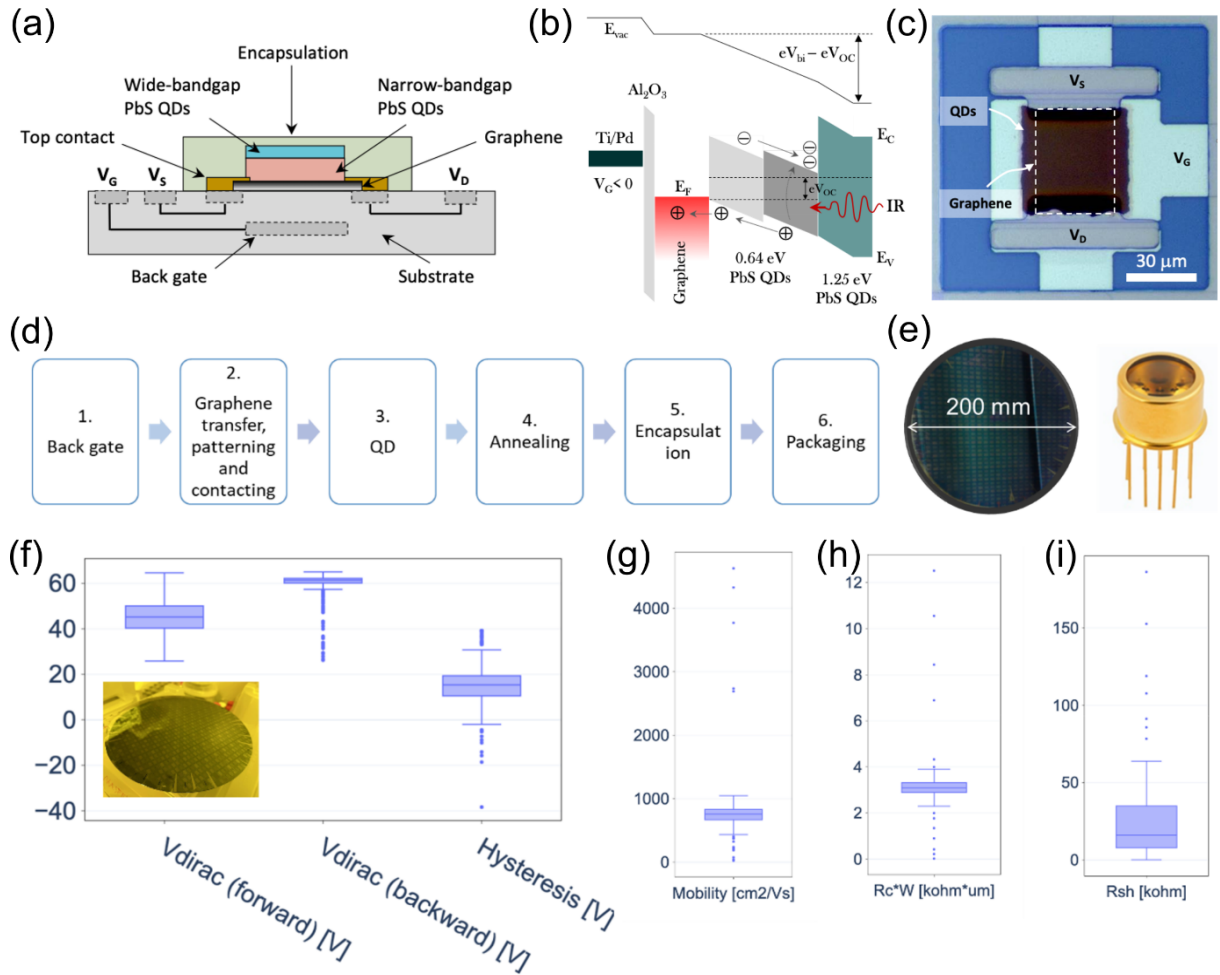

**Figure 1**. (a) Schematic device architecture of the tandem GFET-QD photodetector. (b) Schematic energy band diagram upon illumination. (c) Top-view optical microscope image of tandem GFET-QD photodetector device. (d) Fabrication flow of the GFET-QD photodetector including the key steps of graphene transfer, patterning, contacting, QD deposition, encapsulation, and packaging. (e) As-fabricated 200 mm wafer (left) and later diced, wire-bonded, and packaged single-pixel photodetector in a hermetically sealed semiconductor package (right). (f-i) 200 mm wafer scale statistics on electric metrics of encapsulated GFETs. Box-whisker plots of (f) Dirac points and hysteresis, (g) field effect mobility, (h) contact resistance, and (i) sheet resistance of the GFETs. Inset in (f): a photo of the 200 mm wafer after fabrication.

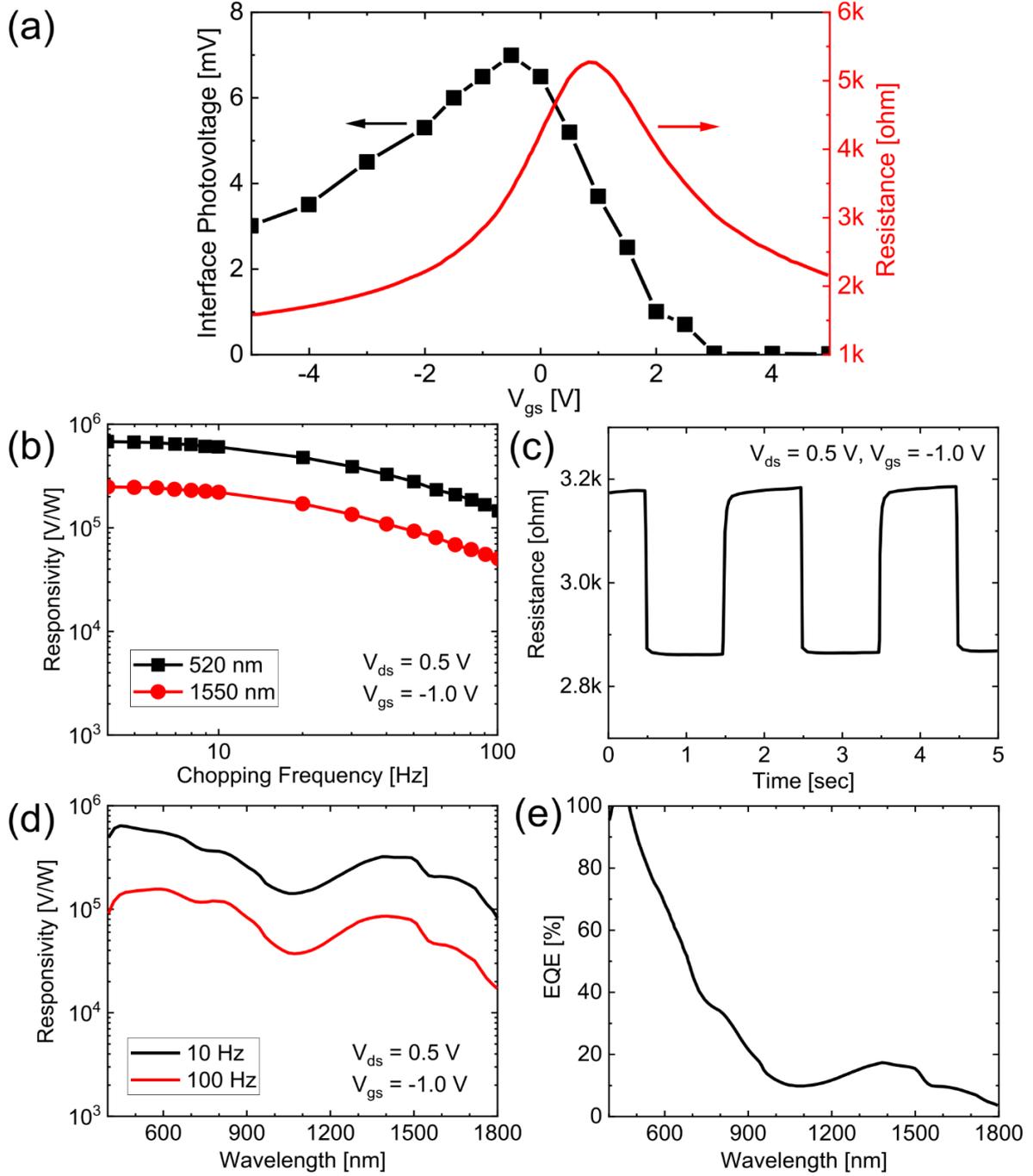

**Figure 2.** Electro-optical response of the GFET-QD photodetector. (a) Photovoltage emerging at the graphene-QD junction shown against the $I_{ds}$-$V_{gs}$ curve of GFET device. (b) Photo-response curves of the photodetector as a function of chopping frequency, with wavelength at 520 and 1550 nm ($V_{ds}$ = 0.5 V, $V_{gs}$ = -1.0 V). (c) Photo-switching behavior of the photodetector towards the pulsed light illumination at a wavelength of 1550 nm, light intensity at 0.7 W/m² $V_{gs}$ = -1.0 V and $V_{ds}$ = 0.5 V. (d) Photo-response curves of the photodetector as a function of wavelength, with frequency at 10 Hz and 100 Hz ($V_{ds}$ = 0.5 V, $V_{gs}$ = -1.0 V). (e) EQE spectral measured at 4 Hz with irradiance level < $10^{-3}$ W/m².

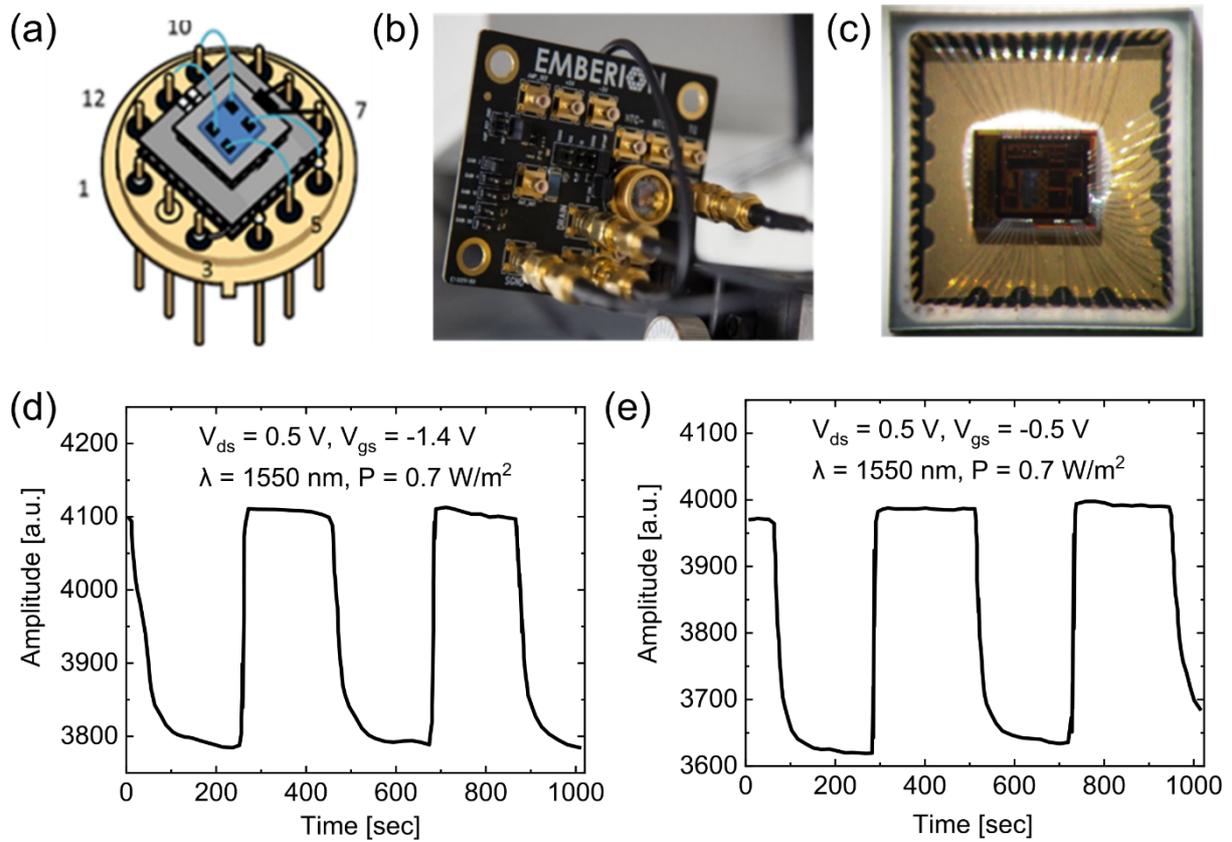

**Figure 3.** The tandem GFET-QD image sensor. (a) Single pixel photo-detector device hermetically sealed into a metal T0-8 semiconductor package with sapphire window. (b) Photograph of the data readout setup via connection to a dedicated readout electronics board. (c) Photograph of a few-pixel GFET array fabricated directly on an ASIC and subsequently wire-bonded and hermetically sealed into a QFN package. Corresponding photo-response of the few-pixel array measured through the ASIC at back gate voltages of $V_{gs}$ = (d) -1.4 V and (e) -0.5 V, $\lambda$ = 1550 nm, $P$ = 0.7 W/m$^2$, $V_{ds}$ = 0.5 V . Units are Arbitrary digital values from the DAC output of the ROIC.

19. Wang, L. *et al.* One-Dimensional Electrical Contact to a Two-Dimensional Material. *Science* **342**, 614–617 (2013).

20. Alexander-Webber, J. A. *et al.* Encapsulation of graphene transistors and vertical device integration by interface engineering with atomic layer deposited oxide. *2D Mater.* **4**, 011008 (2016).

21. Shaygan, M. *et al.* Low Resistive Edge Contacts to CVD-Grown Graphene Using a CMOS Compatible Metal. *ANNALEN DER PHYSIK* **529**, 1600410 (2017).

22. Anzi, L. *et al.* Ultra-low contact resistance in graphene devices at the Dirac point. *2D Mater.* **5**, 025014 (2018).

23. Passi, V. *et al.* Ultralow Specific Contact Resistivity in Metal–Graphene Junctions via Contact Engineering. *Advanced Materials Interfaces* **6**, 1801285 (2019).

24. Compound Quantum Dot–Perovskite Optical Absorbers on Graphene Enhancing Short-Wave Infrared Photodetection | ACS Nano. https://pubs.acs.org/doi/full/10.1021/acsnano.7b00760.

25. 66-4: Invited Paper: Graphene Enhanced QD Image Sensor Technology - Allen - 2021 - SID Symposium Digest of Technical Papers - Wiley Online Library. https://sid.onlinelibrary.wiley.com/doi/abs/10.1002/sdtp.14855.

26. Zhong, H., Zhang, Z., Xu, H., Qiu, C. & Peng, L.-M. Comparison of mobility extraction methods based on field-effect measurements for graphene. *AIP Advances* **5**, 057136 (2015).

27. Chavarin, C. A., Sagade, A. A., Neumaier, D., Bacher, G. & Mertin, W. On the origin of contact resistances in graphene devices fabricated by optical lithography. *Appl. Phys. A* **122**, 58 (2016).

28. Cheng, Z. *et al.* How to report and benchmark emerging field-effect transistors. *Nature Electronics* **5**, 416–423 (2022).

Supporting Information for

# Graphene-Quantum Dot Hybrid Photodetectors from 200 mm Wafer Scale Processing

*Sha Li,[1] Zhenxing Wang,[1,*] Bianca Robertz,[1] Daniel Neumaier,[1,2] Oihana Txoperena,[3] Arantxa Maestre,[3] Amaia Zurutuza,[3] Chris Bower,[4] Ashley Rushton,[4] Yinglin Liu,[4] Chris Harris,[4] Alexander Bessonov,[4] Surama Malik,[4] Mark Allen,[4] Ivonne Medina-Salazar,[4] Tapani Ryhänen,[5] Max C. Lemme[1,6,*]*

[1] AMO GmbH, Otto-Blumenthal-Str. 25, 52074 Aachen, Germany

[2] University of Wuppertal, Chair of Smart Sensor Systems, Lise-Meitner-Str. 13, 42119 Wuppertal, Germany

[3] Graphenea Semiconductor SLU. Donostia, Spain

[4] Emberion Limited, 150-151, Cambridge Science Park, Milton Road, Cambridge, CB4 0GN, UK

[5] Emberion Oy, Metsänneidonkuja 8, 02130 Espoo, Finland

[6] RWTH Aachen University, Chair of Electronic Devices, Otto-Blumenthal-Str. 2, 52074 Aachen, Germany

[*] Correspondence: wang@amo.de, lemme@amo.de

- **Analysis of statistics for 200 mm wafer**

Automated current-voltage measurements were conducted to extract device characteristics and extract performance parameters (drain current $I_{ds}$ vs. gate voltage $V_{gs}$). The measurements were taken under ambient conditions at room temperature, with a source-drain voltage of $V_{ds}$ = 100 mV and $V_{gs}$ swept from -5 to 5 V. A device is categorized as "defective" when the gate leakage ($I_{gs}$) exceeds 100 nA, a short circuit channel ($I_{ds}$ > 5 mA) or an open circuit channel ($I_{ds}$ < 100 nA) is detected, or when the drain current $I_{ds}$ modulation as a function of $V_{gs}$ is < 10%. Out of the 648 devices randomly picked from the 200 mm wafer, 624 were categorized functional and 24 were categorized defective (short or open circuit channels and/or leakage), corresponding to a yield of 96%. Such a high yield demonstrates the successful wafer-scale graphene transfer and includes process steps like the fabrication of graphene contact and gate dielectric deposition between the gate and graphene channel. The fabrication flow was also found to be stable, with reasonable batch-to-batch reproducibility, confirmed by comparable fabrication yield (96% – 98%) and resultant field effect mobility (720-1000 cm$^2$/V·s) from three fabrication runs as summarized in **Table S1**.

The quantified analysis of the graphene electric parameters is based on mobility, doping level, modulation of $I_{ds}$ with $V_{gs}$, as well as contact ($R_c$), and sheet resistance ($R_{sh}$). These were extracted from local back-gated transfer length method (TLM) structures with the channel length L varying from 9 to 89 $\mu$m in steps of 10 $\mu$m and a fixed channel width $W$ of 19 $\mu$m. $R_c$ and $R_{sh}$ were extracted using TLM; mobilities were calculated based on the field-effect mobility model by direct transconductance method (DTM) (see Methods for details on electrical data extraction). The 200 mm wafer scale statistics are summarized in **Figures 1f-1i**, showing histograms of mobility, Dirac point, hysteresis, and $R_c$, $R_{sh}$ of the GFETs. The mobility of the measured devices is 719 ± 172 cm$^2$/V·s. Although higher numbers have been routinely achieved for CVD graphene devices in literature, these values are not critical for the application targeted here. Moreover, the devices are heavily p-doped, which is reflected by the Dirac points centered around $V_{gs}$ ~ 40 V. Possible reasons are the strain and residue introduced by the graphene transfer [1] and fabrication process [2], as well as the fact that the devices were measured in ambient air, which is known to lead to p-doping [3]. We further observed a rather high hysteresis in the transfer characteristics which can be attributed to water molecules that act as charge-trapping centers [4]. The water molecules are likely due to exposure of the as-transferred graphene to ambient air and moisture during the transportation of the wafers between different sites.

- **Analysis of statistics for 150 mm wafer**

For device yield and electric metrics analysis, fully automated transfer characteristics measurement ($I_{ds}$-$V_{gs}$) was performed on the as-fabricated QC wafers under ambient conditions at room temperature, with $V_{ds}$ at 100 mV and $V_{gs}$ in (-5, 5) V. Out of the 440 devices randomly picked from a 150 mm wafer, 433 were measured functional and 7 defective (short or open circuit channels, gate leakage, or low modulation), corresponding to a yield of 98.4%.

For quantified analysis of graphene electric metrics, mobility, doping level, modulation, contact ($R_c$), and sheet resistance ($R_{sh}$) were probed from local back-gated transfer length method (TLM) structures (channel length L varying from 9 to 89 $\mu$m in a step of 10 $\mu$m and channel width $W$ of 19 $\mu$m). $R_c$ and $R_{sh}$ were extracted using TLM; mobility is calculated based on a field-effect mobility model by the direct transconductance method (DTM) (see Methods section for details on electrical data extraction). A summary of graphene-on-wafer quality metrics on 150 mm can be found in **Table S2**, showing mobility, Dirac point, hysteresis, and $R_c$, $R_{sh}$ of the GFETs. The mobility of the measured devices across the whole wafer is 849±351 cm$^2$/V·s, and hysteresis is 0.5 V.

- **Details of the electro-optical measurements**

Detailed electro-optical measurements of as-fabricated 200 mm wafers can be found in **Figure S3**, in general, the transfer curves of the underlying GFETs were good compared to earlier samples, with Dirac point voltage around 1-2 V and low levels of hysteresis, it is not possible to determine if the hysteresis originates in the underlying GFET, or is a result of the absorber stack since the GFET transfer curves cannot be reliability measured before ALD encapsulation. A selection of the GFET-QD transfer curves along with the photo-response measured at $V_{gs}$ = 0 V, $V_{ds}$ = 0.5 V with 24 W/m² of 1550 nm IR light pulsed at 0.5 Hz is shown in **Figure S3b**. Typically, the charge mobility measured for both N and P-branches is approximately 2000 cm²/V·s, with hysteresis of 0.2 V and a Dirac voltage of 1 V. When illuminated with 1550 nm IR light, the average change in channel resistance between dark and illuminated levels is ~ 3.6% with an SNR of approximately 177, with photocurrent of the order of 6 $\mu$A.

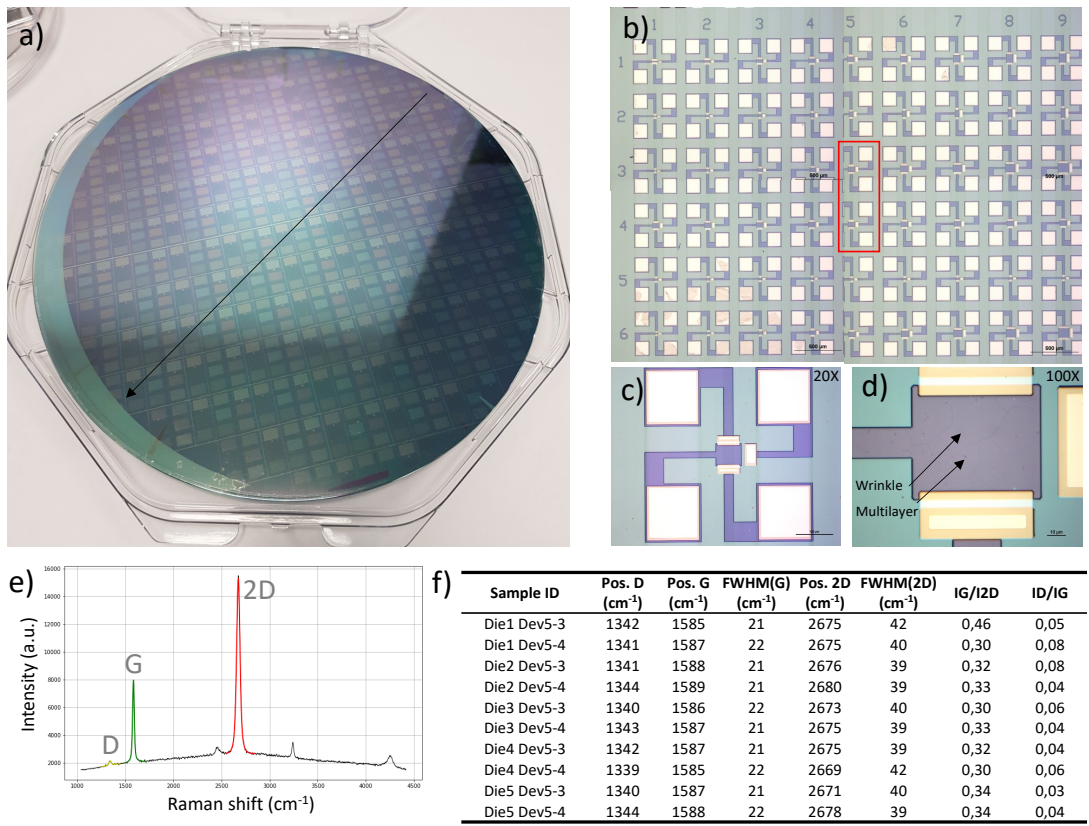

**Figure S1**. (a) 200 mm test wafer, were diagonal dies are routinely inspected. (b) Optical image of a die with a total of 54 devices, where devices No. 5-3 and 5-4 are inspected. (c) 20× and (d) 100× optical images of a representative device. (e) Typical Raman spectrum and f) Raman parameters extracted for a representative graphene-on-wafer sample. Raman spectra were used to assess the graphene quality. The D band is related to disorder and defects in the graphene, while the G band arises from $sp^2$ carbon networks, and the 2D band corresponds to the overtone of the D band.

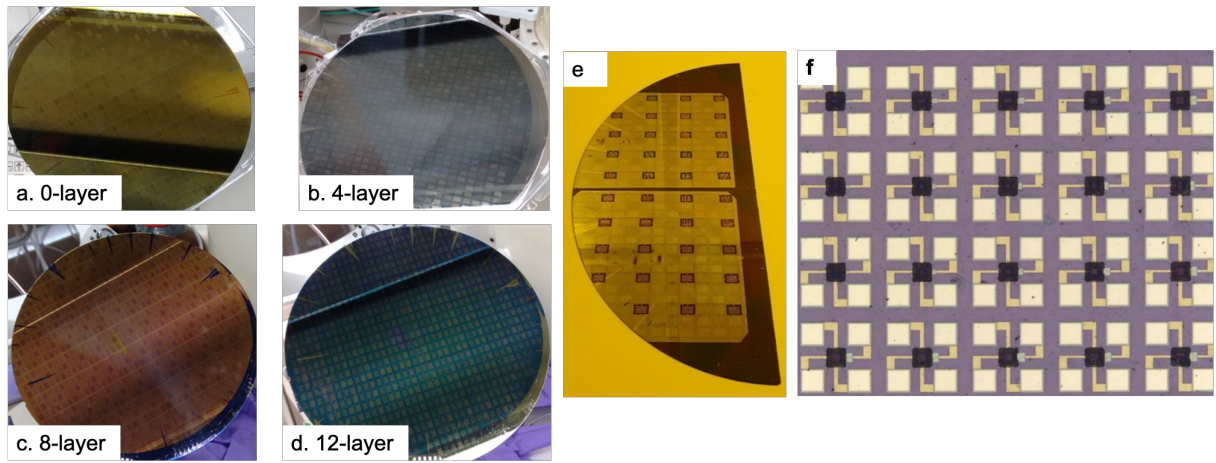

**Figure S2**. Deposition of PbS QD layers on 200 mm Wafer, (a) Wafer with 200 mm graphene layer (b) after deposition of first 4-layers of QDs (c) after 8-layers and (d) after deposition of all 12 layers of the absorber stack. 200 mm wafer after dicing and lithographic patterning and solution etch to remove the absorber layer, except on the graphene channel regions (e). Close-up of device array after patterning of the absorber layer (f).

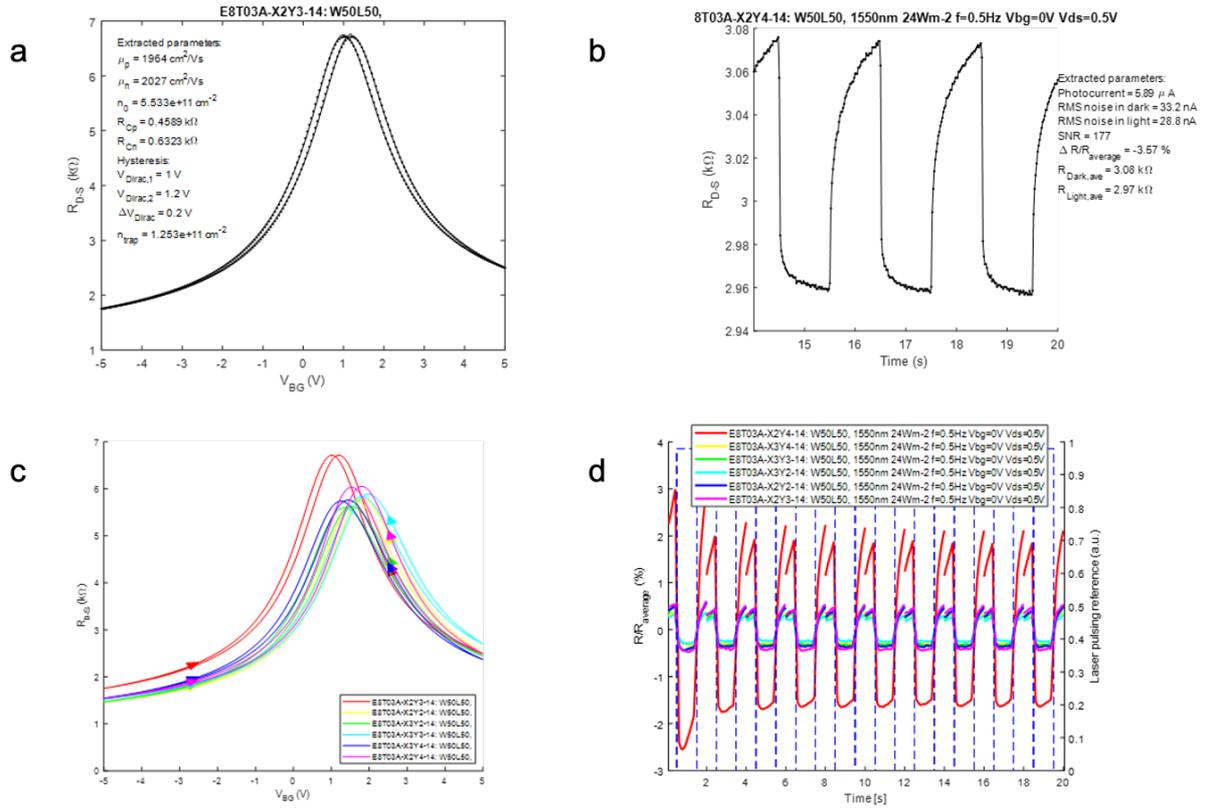

**Figure S3**. (a) Transfer curve for the QD functionalized GFET, showing a Dirac point ~ 1 V with 0.2 V hysteresis. (b) Photo-response at light power of 24 W/m$^2$, 1550 nm light at 0.5 Hz using $V_{gs}$ = 0 V and $V_{ds}$ = 0.5 V. (c) Transfer curves for small array of devices and (d) corresponding photo-response at 1550 nm at light power of 24 W/m$^2$ using $V_{gs}$ = 0 V and $V_{ds}$ = 0.5 V.

**Table S1.** Fabrication results of encapsulated GFETs from different batches.

| Wafer No. | 01 | 02 | 03 |
|---|---|---|---|
| **Wafer scale** | 150 mm | 150 mm | 200 mm |
| **Yield** | 98.4% | 97.5% | 96% |
| **Mobility (cm$^2$/V·s)** | 849 ± 351 | 1055 ± 580 | 719 ± 172 |

**Table S2.** Quality metrics of encapsulated GFETs on 150 mm wafer scale. (*Some of the defective devices fall into multiple non-working categories)

| Wafer No. | 01 |
|---|---|
| **Graphene transfer method** | Semi-dry |
| **Yield** | 433/440 = 98.4%* <br> ○ Gate leakage (I$_{gs}$ > 100 nA): 4 <br> ○ No contact (I$_{ds}$ < 100 nA): 6 <br> ○ Channel short (I$_{ds}$ > 5mA): 0 <br> ○ No modulation (< 10%): 4 |
| **Mobility (cm$^2$/V·s)** | 849 ± 351 |
| **$V_{Dirac}$ for forward scan (V)** | 5.1 ± 0.9 |
| **$V_{Dirac}$ for backward scan (V)** | 5.7 ± 0.9 |
| **$\Delta V_{Dirac}$ (V)** | 0.5 ± 0.7 |
| **$R_c \cdot W$ (kΩ*μm)** | 19 ± 20 |
| **$R_{sh}$ (kΩ/□)** | 6.7 ± 1.5 |
| **Modulation ($I_{ds,max}/I_{ds,min}$)** | 4 ± 0.9 |